\documentclass[twocolumn,showpacs,prc,floatfix]{revtex4}
\usepackage{graphicx}
\usepackage[dvips,usenames]{color}
\usepackage{amsmath}
\newcommand{\p}[1]{(\ref{#1})}
 
\def\beq{\begin{eqnarray}} \def\eeq{\end{eqnarray}}
\def\beqstar{\begin{eqnarray*}} \def\eeqstar{\end{eqnarray*}}
\def\bal{\begin{align}} \def\eal{\end{align}}
\def\beqe{\begin{equation}}\def\eeqe{\end{equation}}

\begin {document}
\title{Unusual temperature behavior of the entropy of \\the antiferromagnetic  spin state  in
nuclear matter \\with an effective finite range interaction}
\author{ A. A. Isayev}
 \affiliation{Kharkov Institute of
Physics and Technology, Academicheskaya Street 1,
 Kharkov, 61108, Ukraine
 }
 \date{\today}
\begin{abstract}The unusual temperature behavior of the entropy of the antiferromagnetic (AFM)
 spin state  in symmetric nuclear matter with the Gogny  D1S interaction, being larger at
 low temperatures  than the entropy  of nonpolarized matter, is related to the
 dependence of the entropy on the effective masses of nucleons in  a spin polarized state.
 The corresponding conditions for comparing the entropies of the AFM
  and nonpolarized states in terms of the effective masses are formulated,
  including the
  low and high temperature limits. It is shown that the unexpected temperature
  behavior of the entropy of the AFM spin state at low   temperatures
  is caused by the violation of   the corresponding low  temperature   criterion.
\end{abstract}
 \maketitle

 The issue of spontaneous appearance of  spin
polarized states in nuclear matter is a topic of a great current
interest due to its relevance in astrophysics. In particular, the
scenarios of supernova explosion and cooling of neutron stars are
essentially different, depending on whether nuclear matter is spin
polarized or not. On the one hand, the models with the effective
Skyrme and Gogny nucleon-nucleon (NN) interaction predict the
occurrence of spin instability in nuclear matter at densities in
the range from $\varrho_0$ to $6\varrho_0$ for different
parametrizations of the NN potential~\cite{R}--\cite{RPV}
($\varrho_0=0.16\,\mbox{fm}^{-3}$ is the nuclear  saturation
density). On the other hand, for the models with the realistic NN
interaction, the ferromagnetic  phase transition seems to be
suppressed up to densities well above
$\varrho_0$~\cite{PGS}--\cite{KS}.

 Here we continue
the research of spin polarizability of nuclear matter with the use
of an effective NN interaction. As was shown in Ref.~\cite{IY2},
in symmetric nuclear matter with the Gogny D1S effective
interaction
 the antiferromagnetic (AFM) spin ordering with the oppositely
directed neutron and proton spins sets in, beginning from some
critical density (at zero temperature,
$\varrho_c\approx3.8\varrho_0$). At finite  temperature, the
entropy of the AFM spin state demonstrates the unusual behavior
being larger than the entropy of the nonpolarized state at low
enough temperatures. The main goal of this work is to clarify this
unexpected moment, utilizing the approximation of the effective
mass in the single particle spectrum of nucleons. As will be shown
later, this approximation is quite successful in reproducing the
entropy of a spin polarized state for all relevant temperatures.
The use of low and high temperature expressions for the entropy
will  allow us to get the corresponding conditions in terms of the
effective masses for comparing the entropies of spin polarized and
nonpolarized states.

 Now we stop on the basic equations of the theory~\cite{I,IY}.
   Given the
possibility of phase transitions to the states with parallel and
antiparallel ordering of neutron and proton spins, the
distribution function $f$
 can be expanded in the
Pauli matrices $\sigma_i$ and $\tau_k$ in spin and isospin
spaces
\begin{align} f({\bf p})&= f_{00}({\bf
p})\sigma_0\tau_0+f_{30}({\bf p})\sigma_3\tau_0\label{7.2}\\
&\quad + f_{03}({\bf p})\sigma_0\tau_3+f_{33}({\bf
p})\sigma_3\tau_3. \nonumber  \end{align} Expressions for  the
distribution functions $f_{00},f_{30},f_{03},f_{33}$
  read~\cite{I,IY}
\begin{align}
f_{00}&=\frac{1}{4}\{n(\omega_{n\uparrow})+n(\omega_{p\uparrow})+n(\omega_{n\downarrow})
+n(\omega_{p\downarrow}) \},\nonumber
 \\
f_{30}&=\frac{1}{4}\{n(\omega_{n\uparrow})+n(\omega_{p\uparrow})-n(\omega_{n\downarrow})-
n(\omega_{p\downarrow})
\},\label{2.4}\\
f_{03}&=\frac{1}{4}\{n(\omega_{n\uparrow})-n(\omega_{p\uparrow})+n(\omega_{n\downarrow})-
n(\omega_{p\downarrow})
\},\nonumber\\
f_{33}&=\frac{1}{4}\{n(\omega_{n\uparrow})-n(\omega_{p\uparrow})-n(\omega_{n\downarrow})+
n(\omega_{p\downarrow}) \}.\nonumber
 \end{align} Here $n(\omega)=\{\exp(\omega/T)+1\}^{-1}$ and
\begin{align}
\omega_{n\uparrow}&=\varepsilon_{0}+\tilde\varepsilon_{00}+\tilde\varepsilon_{30}+
\tilde\varepsilon_{03}+
\tilde\varepsilon_{33}-\mu_{n\uparrow},\;\nonumber\\
\omega_{p\uparrow}&=\varepsilon_{0}+\tilde\varepsilon_{00}+\tilde\varepsilon_{30}-
\tilde\varepsilon_{03}-\tilde\varepsilon_{33}-\mu_{p\uparrow},\;\label{2.5}\\
\omega_{n\downarrow}&=\varepsilon_{0}+\tilde\varepsilon_{00}-\tilde\varepsilon_{30}+
\tilde\varepsilon_{03}-\tilde\varepsilon_{33}-\mu_{n\downarrow},\;\nonumber\\
\omega_{p\downarrow}&=\varepsilon_{0}+\tilde\varepsilon_{00}-\tilde\varepsilon_{30}-
\tilde\varepsilon_{03}+\tilde\varepsilon_{33}-\mu_{p\downarrow},\;\nonumber\end{align}
are the branches
 of the quasiparticle spectrum corresponding to neutrons and protons with spin up and spin down, and
$\mu_{\tau\sigma}$ are their respective chemical potentials
($\tau=n,p$; $\sigma= \uparrow,\downarrow$). Under derivation of
Eqs.~\p{2.4}, \p{2.5}, it is  assumed
 that the populations of  neutrons and protons with spin up and spin down are held fixed.
In Eq.~\p{2.5}, $\varepsilon_0({\bf p})$ is the free single
particle spectrum,  and
$\tilde\varepsilon_{00},\tilde\varepsilon_{30},\tilde\varepsilon_{03},\tilde\varepsilon_{33}$
are the Fermi liquid (FL) corrections to the free single particle
spectrum, related to the  FL amplitudes $U_0({\bf k}),...,U_3({\bf
k}) $ by formulas \begin{align}
 \quad\quad\quad\quad\quad\quad
 \tilde\varepsilon_{00}({\bf
p})&=\frac{1}{2\cal V}\sum_{\bf q}U_0({\bf k})f_{00}({\bf
q}),\;{\bf k}=\frac{{\bf p}-{\bf q}}{2}, \nonumber  
\\
 \tilde\varepsilon_{30}({\bf
p})&=\frac{1}{2\cal V}\sum_{\bf q}U_1({\bf k})f_{30}({\bf
q}), \label{14.1} \\ 
\tilde\varepsilon_{03}({\bf p})&=\frac{1}{2\cal V}\sum_{\bf
q}U_2({\bf k})f_{03}({\bf q}), \nonumber\\ 
  \tilde\varepsilon_{33}({\bf
p})&=\frac{1}{2\cal V}\sum_{\bf q}U_3({\bf k})f_{33}({\bf q}).
\nonumber
\end{align}
 The distribution functions $f_{00},f_{03},f_{30},f_{33}$, in
turn, should  satisfy the normalization conditions for the total
density $\varrho_n+\varrho_p=\varrho$, excess of neutrons over
protons $\varrho_n-\varrho_p\equiv\alpha\varrho$, ferromagnetic
(FM)
$\varrho_\uparrow-\varrho_\downarrow\equiv\Delta\varrho_{\uparrow\uparrow}$
and antiferromagnetic (AFM)
$(\varrho_{n\uparrow}+\varrho_{p\downarrow})-
(\varrho_{n\downarrow}+\varrho_{p\uparrow})\equiv\Delta\varrho_{\uparrow\downarrow}$
spin order parameters, respectively  ($\alpha$ being the isospin
asymmetry parameter,
$\varrho_\uparrow=\varrho_{n\uparrow}+\varrho_{p\uparrow}$ and
$\varrho_\downarrow=\varrho_{n\downarrow}+\varrho_{p\downarrow}$
with  $\varrho_{n\uparrow}, \varrho_{n\downarrow}$
 and
 $\varrho_{p\uparrow},\varrho_{p\downarrow}$ being the neutron and
 proton number densities with spin up and spin down, respectively).  To check
the thermodynamic stability of different solutions of the
self-consistent equations~\p{2.4}--\p{14.1}, it is necessary to
compare the corresponding free energies $F=E-TS$, where $E$ is the
energy functional and the entropy reads \bal S&=-\sum_{\bf
p}\sum_{\tau=n,\,p}\,\sum_{\sigma=\uparrow,\,\downarrow}\{n(\omega_{\tau\sigma})\ln
n(\omega_{\tau\sigma})\label{entr}\\ &\quad+\bar
n(\omega_{\tau\sigma})\ln \bar n(\omega_{\tau\sigma})\}, \;\bar
n(\omega)=1-n(\omega).\nonumber
\end{align}

The single particle energies~\p{2.5} have the following general
structure \beqe \omega_{\tau\sigma}({\bf
k})=\omega_{\tau\sigma}^0({\bf k})+U_{\tau\sigma}({\bf k}),\quad
\omega_{\tau\sigma}^0({\bf k})\equiv \frac{\hbar^2{\bf
k}^2}{2m_{0}}-\mu_{\tau\sigma}, \eeqe where $m_0$ is the bare
nucleon mass, $U_{\tau\sigma}$ is the single particle potential.
Its momentum dependence can be characterized by the effective mass
$m_{\tau\sigma}(k)$, defined as \beqe
\frac{m_0}{m_{\tau\sigma}(k)}=1+\frac{m_0}{\hbar^2k}\frac{dU_{\tau\sigma}(k)}{dk}.\label{efmass}\eeqe
If to use  the quadratic approximation for the single particle
potential
$$U_{\tau\sigma}({\bf k})\approx U_{\tau\sigma}(0)+\biggl(\frac {1}{2k}
\frac {dU_{\tau\sigma}(k)} {dk}\biggr)\biggr|_
{k=k_{F_{\tau\sigma}}}\cdot k^2,$$ where $k_{F_{\tau\sigma}}$ is
the Fermi momentum of nucleons in the state ($\tau,\sigma$), then
the single particle energy can be represented in the form \beqe
\omega_{\tau\sigma}({\bf k})=\frac{\hbar^2{\bf
k}^2}{2m_{\tau\sigma}}+U_{\tau\sigma}(0)-\mu_{\tau\sigma},\;
m_{\tau\sigma}\equiv m_{\tau\sigma}(k_{F_{\tau\sigma}}).\label{7}
\eeqe Within this approximation, all thermodynamic quantities can
be easily calculated, analogously to the case of a free Fermi gas.
Note that in order to get the effective mass $m_{\tau\sigma}$,  it
is necessary to find the explicit single particle potential as a
result of the solution of the self-consistent
equations~\p{2.4}--\p{14.1}.

  Further we will consider   symmetric nuclear
matter with the  Gogny D1S interaction as a potential of NN
interaction. It was shown in Ref.~\cite{IY2} that in this case
 the AFM spin ordering can be realized only among the states with
the collinear spin ordering. In the AFM spin state,
$\varrho_{n\uparrow}=\varrho_{p\downarrow},
\varrho_{n\downarrow}=\varrho_{p\uparrow}$, and, hence,
$\mu_{n\uparrow}=\mu_{p\downarrow},
\mu_{n\downarrow}=\mu_{p\uparrow}$. Besides, we have only two
different branches in the quasiparticle spectrum,
$\omega_{n\uparrow}=\omega_{p\downarrow}$, and
$\omega_{n\downarrow}= \omega_{p\uparrow}$, and, as a consequence,
only two different effective masses,
$m_{n\uparrow}=m_{p\downarrow}$, and $m_{n\downarrow}=
m_{p\uparrow}$.
  The neutron $\Pi_n= \frac{\varrho_{n\uparrow}-\varrho_{n\downarrow}}{\varrho_n}$
and proton $\Pi_p=\frac{\varrho_{p\uparrow}-\varrho_{p\downarrow}}{\varrho_p}$ spin
polarization parameters for the AFM spin state are
opposite in sign and equal to
$$\Pi_n=-\Pi_p=\frac{\Delta\varrho_{\uparrow\downarrow}}{\varrho}\equiv \Pi.$$

\begin{figure}[tb] 
\begin{center}
\includegraphics[height=7.0cm,width=8.6cm,trim=48mm 144mm 56mm 66mm,
draft=false,clip]{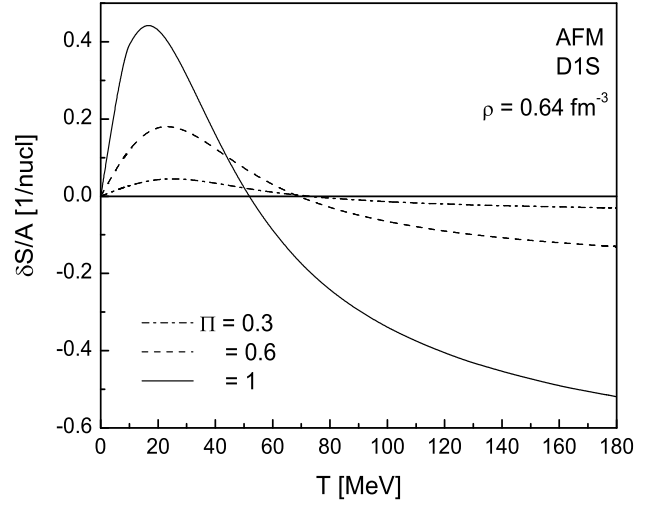}\end{center} \caption{The entropy per
nucleon, measured from its value in the nonpolarized state, for
the AFM spin state  as a function of temperature at different
polarizations.} \label{fig1}
\end{figure}

In Fig.~\ref{fig1}, the difference between the entropies per
nucleon of the AFM and nonpolarized states is shown as a function
of temperature at different fixed polarizations. The value $\Pi=1$
corresponds to the totally AFM polarized nuclear matter. One can
see that for low temperatures the entropy of the AFM state  is
larger than the entropy of the normal state. It looks like the AFM
state at low finite temperatures is less ordered than the
nonpolarized state. Under a further increase of temperature the
difference between the entropies changes the sign and becomes
negative, that corresponds to the intuitively expected behavior.

\begin{figure}[tb] 
\begin{center}
\includegraphics[height=7.0cm,width=8.6cm,trim=48mm 144mm 56mm 66mm,
draft=false,clip]{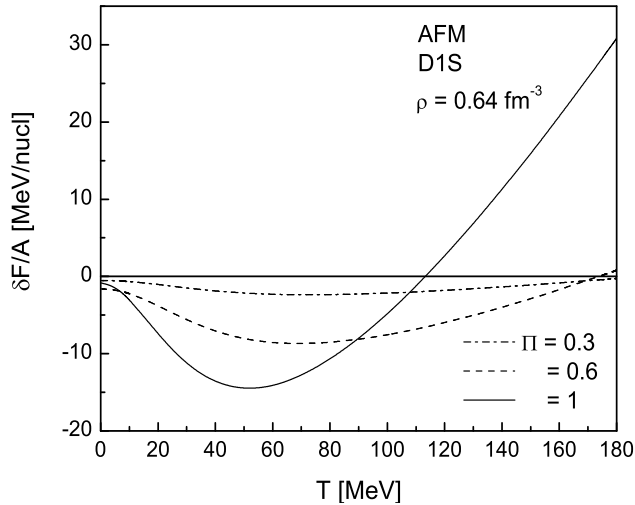}\end{center} \caption{Same as in
Fig.~\ref{fig1}, but for the free energy per nucleon, measured
from its value in the nonpolarized state.} \label{fig1a}
\end{figure}
Fig.~\ref{fig1a} shows the difference between the free energies
per nucleon of the AFM and nonpolarized states as a function of
temperature for the same polarizations as in Fig.~\ref{fig1}. In
contrast to the difference between the entropies, the difference
between the free energies preserves its negative sign for  all
temperatures below the corresponding critical temperature,  and,
hence, the AFM spin state is thermodynamically preferable as
compared to  the nonpolarized state for the whole corresponding
temperature domain. Therefore, the unusual temperature behavior of
the entropy of the AFM spin state doesn't lead to the instability
of the polarized state at low temperatures. Note that if to assume
the quadratic approximation for the dependence of the free energy
per nucleon of the polarized state on the spin polarization
parameter~\cite{VB},
$$\frac{F(\varrho,T,\Pi)}{A}=\frac{F(\varrho,T,\Pi=0)}{A}+\gamma(\varrho,T)\Pi^2,$$
then the curve in Fig.~\ref{fig1a}, corresponding to $\Pi=1$, in
fact, shows the  dependence of the spin-isospin symmetry parameter
$\gamma$ on temperature. Its negative value proves the stability
of the AFM spin state at temperatures below the critical
temperature, as clarified above.

\begin{figure}[tb] 
\begin{center}
\includegraphics[height=7.0cm,width=8.6cm,trim=51mm 146mm 49mm 66mm,
draft=false,clip]{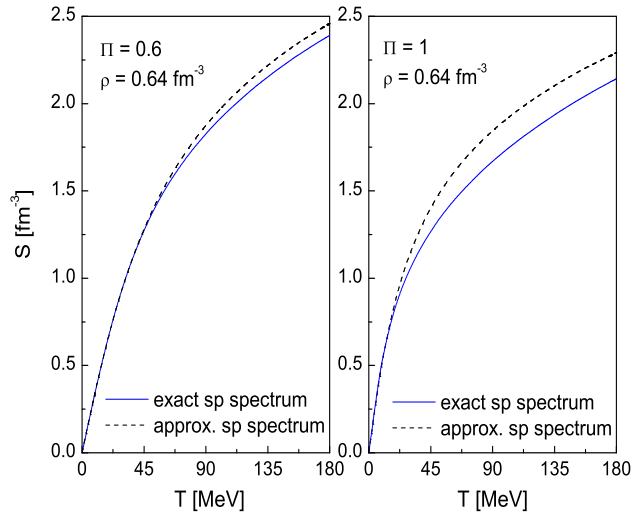}
\end{center}
\caption{(Color online) The density of entropy for the AFM spin
state as a function of temperature, calculated with the exact
(solid line) and approximated (dashed line) single particle
energies at the spin polarization parameter $\Pi=0.6$ (left) and
$\Pi=1$ (right).} \label{fig3}
\end{figure}

\begin{figure}[tb] 
\begin{center}
\includegraphics[height=7.0cm,width=8.6cm,trim=48mm 144mm 54mm 66mm,
draft=false,clip]{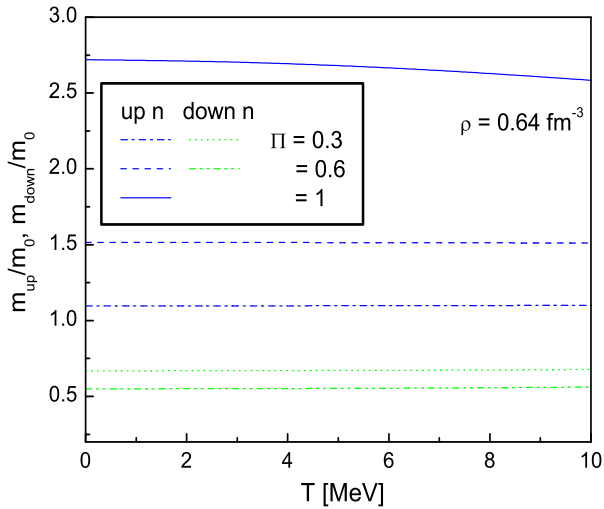} 
\end{center}
\caption{(Color online) The effective masses of spin up and spin
down neutrons in units of the bare nucleon mass as functions of
temperature in the low temperature domain at different
polarizations.} \label{fig4}
\end{figure}

To understand the unusual behavior of the entropy of the AFM spin
state at low temperatures, we utilize the quadratic approximation
for the single particle spectrum of nucleons,
 Eq.~\p{7}. Note that, adopting this approximation, we self-consistently determine also
  the chemical potentials to guarantee the fulfillment of the normalization conditions
  for the distribution functions.    The results of the numerical
 determination of the entropy density, based on the exact and
 approximated forms of the single particle energies, are compared in
 Fig.~\ref{fig3}. It is seen that the approximation turns out to be quite satisfactory,
 especially in the region of low temperatures, and the
 disagreement does not exceed $7\%$ even for the less favorable case
 of totally polarized matter in the entire temperature domain under consideration.  Therefore, we
 can provide the low temperature expansion for the entropy using
 the approximation of the effective mass in the single particle
 energies, as described in detail, e.g., in Ref.~\cite{LL}. Then, requiring for
 the difference between the entropies
 per nucleon of the AFM and nonpolarized states to be negative, one can get
 the condition
\beqe D_1\equiv \frac{m_{n\uparrow}}{m^*}(1+\Pi)^\frac{1}{3}+
\frac{m_{n\downarrow}}{m^*}(1-\Pi)^\frac{1}{3}-2<0.\label{9}\eeqe
Here $m^*$ is the effective mass of a nucleon in nonpolarized
nuclear matter at the corresponding temperature and density. The
low temperature condition~\p{9} is valid until
$T/\varepsilon_{F_{n\sigma}}\ll1$,
$\varepsilon_{F_{n\sigma}}=\frac{\hbar^2k^2_{F_{n\sigma}}}{2m_{n\sigma}}$
being the Fermi energy of neutrons with spin up
($\sigma=\uparrow$) and spin down ($\sigma=\downarrow$). The
calculations show that at the given density ($\varrho=0.64\,{\rm
fm}^{-3}$) and polarizations, the corresponding temperature
interval extends approximately up to $T=10\,{\rm MeV}$. Besides,
under derivation of the condition~\p{9} it is assumed that the
effective masses are temperature independent. Fig.~\ref{fig4}
shows the dependence of the effective masses
$m_{n\uparrow},m_{n\downarrow}$ on temperature in the temperature
domain, where the low temperature expansion holds true. It is seen
that  the effective masses for partially AFM polarized nuclear
matter at the given polarizations are practically independent on
temperature, and for totally polarized matter the change in the
effective mass is about $5\%$ in this temperature domain.
Therefore, the use of Ineq.~\p{9} is quite justified for comparing
the entropies of the AFM and nonpolarized states at low
temperatures.

\begin{figure}[tb] 
\begin{center}
\includegraphics[height=7.0cm,width=8.6cm,trim=51mm 146mm 49mm 66mm,
draft=false,clip]{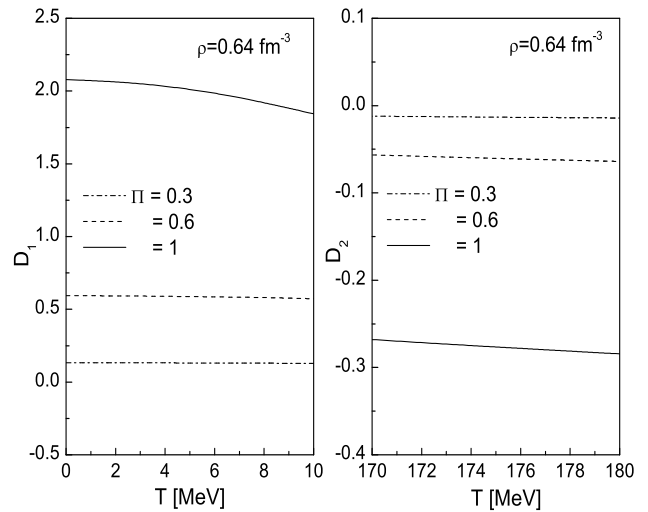} 
\end{center}
\caption{The differences  $D_1$ (left) and $D_2$ (right) in
Ineqs.~\p{9} and \p{D2} as functions of temperature  at different
polarizations.} \label{fig5}
\end{figure}

Fig.~\ref{fig5} shows the dependence of the l.h.s. $D_1$ of
Ineq.~\p{9} on temperature in the low temperature domain under
consideration at different polarizations. It is seen that the
quantity $D_1$ is positive and increases with the spin
polarization. This explains the unexpected behavior of the entropy
of the AFM spin state, being larger than that of the nonpolarized
state at low temperatures.

Note that at higher temperatures the entropy of the AFM spin state
becomes smaller than the entropy of the nonpolarized state. To
explain this, we can again use  the  approximation of the
effective mass in the single particle energies for getting the
high temperature expression for the entropy. If
$\varrho_{n\sigma}\lambda_{n\sigma}^3\ll 1$
($\lambda_{n\sigma}=\sqrt{\frac{2\pi\hbar^2}{m_{n\sigma}T}}$ is
the thermal wavelength of neutrons with spin up and spin down),
then the condition for the difference between the entropies per
nucleon of the AFM and nonpolarized states to be negative is \beqe
D_2\equiv\biggl(\frac{m_{n\uparrow}^\frac{3}{2}}{1+\Pi}\biggl)^{\frac{1+\Pi}{2}}
\biggl(\frac{m_{n\downarrow}^\frac{3}{2}}{1-\Pi}\biggl)^{\frac{1-\Pi}{2}}\frac{1}
{m^{*\frac{3}{2}}}-1<0.\label{D2}\eeqe
 Fig.~\ref{fig5} shows the dependence of the quantity $D_2$ on
temperature in the high temperature region at the given density
and polarizations. One can see that the condition~\p{D2} is
fulfilled, and, hence, the entropy of the AFM spin state turns out
to be smaller than that of the nonpolarized state. Note that from
the derivation procedure of the high temperature expression for
the entropy it follows that the effective masses in
 Ineq.~\p{D2} can be temperature dependent.

Thus, using the approximation of the effective mass, it is
possible to explain both the low temperature and high temperature
peculiarities of the entropy of spin polarized nuclear matter with
the D1S Gogny interaction. In Ref.~\cite{RPV}, it was found that
the entropy of spin polarized neutron matter with the Skyrme
effective interaction above some critical density $\varrho_{S}$ is
larger than that of nonpolarized matter. As a consequence, the
critical density for the appearance of a spin polarized state
decreases with temperature, contrary to the intuitive suggestion.
However, there is an important difference between the cases with
the finite range Gogny and zero range Skyrme forces: While for the
Skyrme interaction the effective masses in Eq.~\p{efmass} are
momentum and temperature independent, for the Gogny interaction
they do depend on momentum and temperature. By this reason, the
difference between the entropies of polarized and nonpolarized
states is positive for all temperatures at densities above
$\varrho_S$ for the Skyrme interaction. For the Gogny interaction,
 this difference changes the sign from the positive one at low
 temperatures to the negative one at high temperatures. The critical
 density $\varrho_S$, above which the difference of the entropies
 becomes positive at low temperatures, depends on polarization,
 for example,
 $\varrho_S(\Pi=0.3)\approx0.07\,\rm{fm}^{-3}$,
 $\varrho_S(\Pi=0.6)\approx0.08\,\rm{fm}^{-3}$, and
 $\varrho_S(\Pi=1)\approx0.17\,\rm{fm}^{-3}$. For comparison, for totally polarized neutron matter
 with the
 SLy4 Skyrme interaction $\varrho_S\approx0.15\,\rm{fm}^{-3}$~\cite{RPV}. Since the entropy
 of the AFM spin state  is  smaller
 than that of nonpolarized matter at high temperatures,  the critical density
 for the appearance of the AFM  state increases with temperature~\cite{I2}, in agreement
 with the intuitive considerations and contrary to the scenario with the Skyrme
 interaction~\cite{RPV}.

 In summary, it has been shown that the entropy of the AFM spin state
 in symmetric nuclear matter with the Gogny  D1S interaction demonstrates the unusual
 behavior, being larger at low temperatures  than the entropy  of nonpolarized matter.
 By comparing the free energies of polarized and
 nonpolarized  states, it has been clarified that this unconventional
  temperature behavior doesn't lead to the  instability of the AFM state. This entropy peculiarity
    has been  related to the
 dependence of the entropy on the effective masses of nucleons in
 a spin polarized state, which for the finite range D1S Gogny interaction
 do
 depend on temperature. The corresponding conditions for comparing
 the entropies of the AFM
  and nonpolarized states in terms of the effective masses have been formulated, including
  the low and high temperature limits. It has been shown that the
  unexpected temperature behavior of the entropy at low temperatures is caused by the violation of
  the corresponding low temperature criterion.

\end{document}